\newcommand{\et}{\hspace{-0.08in}{\bf .}\hspace{0.1in}}
\newcommand{\BOX}{\hbox {$\sqcap$ \kern -1em $\sqcup$}}

\renewcommand{\to}{\rightarrow}
\newcommand{\tensor}{\otimes}
\newcommand{\maps}{\colon}

\newcommand{\iso}{\cong}

\renewcommand{\S}{{\rm S}}
\newcommand{\G}{{\cal G}}
\newcommand{\U}{{\rm U}}
\newcommand{\SU}{{\rm SU}}
\newcommand{\SO}{{\rm SO}}
\newcommand{\OO}{{\rm O}}

\newcommand{\Spin}{{\rm Spin}}

\newcommand{\R}{{\mathbb R}}
\newcommand{\C}{{\mathbb C}}

\newtheorem{prop}{Proposition}
\newtheorem{cor}{Corollary}

\documentclass{article}
\usepackage{amsfonts,amssymb}

\title{Calabi--Yau Manifolds and the Standard Model}
   \author{John C.\ Baez\\
   Department of Mathematics,  University of California\\
   Riverside, California 92521 \\
   USA \\
   \medskip
   email: baez@math.ucr.edu \\}
   \date{November 7, 2005}
\begin{document}
\bibliographystyle{plain}
\maketitle

\begin{abstract}
\noindent
For any subgroup $G \subseteq \OO(n)$,
define a \textbf{{\textit G}-manifold} 
to be an $n$-dimensional Riemannian manifold 
whose holonomy group is contained in $G$.  Then a $G$-manifold
where $G$ is the Standard Model gauge group 
is precisely a Calabi--Yau manifold of 10 real dimensions whose 
tangent spaces split into orthogonal 4- and 6-dimensional subspaces, 
each preserved by the complex structure and parallel transport.
In particular, the product 
of Calabi--Yau manifolds of dimensions 4 and 6 gives such
a $G$-manifold.  Moreover, any such $G$-manifold 
is naturally a spin manifold, and Dirac spinors on 
this manifold transform in the representation of $G$ corresponding to 
one generation of Standard Model fermions and their antiparticles.
\end{abstract}

\vskip 2em

\noindent
The purpose of this note is to point out a curious relation
between the mathematics of the Standard Model and the geometry of 
Calabi--Yau manifolds.   
The gauge group of the Standard Model is often said to be
$\SU(3) \times \SU(2) \times \U(1)$, but it is well known
that a smaller group is sufficient.  The reason is that 
$\SU(3) \times \SU(2) \times \U(1)$ has a normal subgroup $N$
that acts trivially on all the particles in the Standard Model.  
This subgroup has 6 elements, and it is generated by 
\[      (e^{2\pi i/3}I, -I, e^{\pi i/3}) \in 
\SU(3) \times \SU(2) \times \U(1) .\]
The resulting quotient 
\[         \G = (\SU(3) \times \SU(2) \times \U(1))/N   \] 
could be called the `true' gauge group of the Standard Model.

There is an isomorphism
\[   
\begin{array}{ccl}
  \G &\iso& \S(\U(3) \times \U(2))  \\
     &:=&
\{ x \in \SU(5) :
x = \left( \begin{array}{cc} 
  g & 0 \\
  0 & h 
     \end{array} \right)  , \;
g \in \U(3), \; h \in \U(2) \}  
\end{array}
\]
since there is an onto homomorphism
\[  
\begin{array}{ccl}
\SU(3) \times \SU(2) \times \U(1) &\to& \S(\U(3) \times \U(2)) \\ 
\\
 (g,h,\alpha) &\mapsto&   
 \left( \begin{array}{cc} 
 \alpha^{-2} g & 0 \\
  0 & \alpha^{3} h 
\end{array} \right)
\end{array}
\]
whose kernel is precisely $N$.  The Georgi--Glashow
$\SU(5)$ grand unified theory \cite{GG} relies on this fact, since
$\S(\U(3) \times \U(2))$  is a subgroup of $\SU(5)$ in an obvious way,
while $\SU(3) \times \SU(2) \times \U(1)$ is {\it not} a subgroup
of $\SU(5)$.  The $\SO(10)$ grand unified theory also depends on 
this fact, since it uses the inclusions 
\[ \G \iso \S(\U(3) \times \U(2)) \hookrightarrow \SU(5) 
\hookrightarrow \SO(10) .\]

The beauty of the $\SO(10)$ theory is that the fermions and
antifermions in one generation of the Standard Model form the 
Dirac spinor representation of $\SO(10)$ --- or more precisely, of its 
double cover $\Spin(10)$.  While this spinor representation
is not single-valued on $\SO(10)$, it becomes
so when restricted to the simply-connected subgroup $\SU(5)$.
It is then equivalent to the 
natural representation of $\SU(5)$ on the exterior algebra
$\Lambda \C^5$.  When restricted further to $\G = S(\U(3) \times \U(2))$,
it gives the representation of $\G$ on the fermions and antifermions
in one generation of the Standard Model, as follows:
\[
\begin{array}{ccrl}
\Lambda^0(\C^5) &\iso&       \C & \ni \overline{\nu}_L  \\ 
\\
\Lambda^1(\C^5) &\iso&       \C^3           
                            & \ni d^r_R, d^b_R, d^g_R  \\
                &    &        \oplus \quad \C^2           
                            & \ni \overline{e}_R, \overline{\nu}_R \\
\\
\Lambda^2(\C^5) &\iso&        \Lambda^2 \C^3    
                            & \ni \overline{u^r}_L, \overline{u^g}_L,
                                  \overline{u^b}_L    \\
                &    &         \oplus \quad \C^3 \tensor \C^2 
                            & \ni u^r_L, u^g_L, u^b_L, d^r_L, d^g_L, d^b_L  \\   
                &    &        \oplus \quad \Lambda^2 \C^2    
                            & \ni \overline{e}_L     \\
\\
\Lambda^3(\C^5) &\iso&       (\Lambda^2 \C^3)^*    
                            & \ni u^r_R, u^g_R, u^b_R    \\
                &     &       \oplus \quad (\C^3 \tensor \C^2)^* 
                            & \ni \overline{u^r}_R, \overline{u^g}_R, 
                                  \overline{u^b}_R, \overline{d^r}_R, 
                                  \overline{d^g}_R, \overline{d^b}_R  \\ 
                &    &        \oplus \quad (\Lambda^2 \C^2)^*    
                            & \ni   e_R   \\
\\
\Lambda^4(\C^5) &\iso&(\C^3)^*      
                            & \ni \overline{d^r}_L, \overline{d^g}_L,
                                  \overline{d^b}_L            \\
                &    &\oplus \quad (\C^2)^*           
                     & \ni e_L, \nu_L \\
\\
\Lambda^5(\C^5) &\iso&        \C 
                            & \ni \nu_R  
\end{array}
\]
Here $\C^3$ and $\C^2$ stand for the fundamental representations of
$\SU(3)$ and $\SU(2)$, respectively; $R$ and $L$ stand for right-
and left-handed states, while $r,g,b$ denote quark colors.  The 
conjugate-linear Hodge star operator on $\Lambda \C^5$ maps particles 
to their antiparticles.  For more details, see for example \cite{Zee}.

We can relate these facts to the geometry of Calabi--Yau manifolds
as follows. For any subgroup $G \subseteq \OO(n)$, let us say that
an $n$-dimensional Riemannian manifold $X$ is a \textbf{\textit{G}-manifold}
if the holonomy group at any point $x \in X$ is contained in $G$ \cite{Joyce}.
Some examples are familiar:
\begin{itemize}
\item A $\U(n/2)$-manifold is a K\"ahler manifold.
\item An $\SU(n/2)$-manifold is a Calabi--Yau manifold.
\end{itemize}
Here we assume $n$ is even and identify $\R^{n}$ with $\C^{n/2}$ to 
see $\U(n/2)$ as a subgroup of $\OO(n)$.  In what follows when we
speak of the dimension of a Calabi--Yau manifold we mean its 
real dimension, namely $n$.  

If $G$ is the group of linear transformations of $\R^n$ that 
preserve some structure, 
each tangent space of a $G$-manifold will acquire this structure, 
and this structure will be preserved by parallel transport with respect
to the Levi--Civita connection.  In the case of a K\"ahler 
manifold, this structure is a complex structure 
$J_x \maps T_xX \to T_xX$ compatible with the Riemannian metric.
This makes each tangent space into a complex inner product space.
In the case of a Calabi--Yau manifold there is also some additional
structure.

Which manifolds are $\G$-manifolds
when $\G = \S(\U(3) \times \U(2))$ is the true gauge group of the Standard
Model, regarded as a subgroup of $\OO(10)$ as above?
The answer is simple:

\begin{prop} \et    A $\G$-manifold is the same as 
a 10-dimensional Calabi--Yau manifold $X$ each of whose tangent
spaces is equipped with a splitting
into orthogonal 4- and 6-dimensional subspaces
that are preserved both by the complex structure $J_x \maps T_xX
\to T_xX$ and by parallel transport.
\end{prop}

The proof of this fact is immediate, since $\G$ 
is precisely the subgroup of $\SU(5) \subseteq \OO(10)$ that preserves 
a splitting of $\R^{10} \iso \C^5$ into orthogonal 
2- and 3-dimensional complex subspaces.
 
\begin{cor} \et If $M$ and $K$ are Calabi--Yau manifolds of 
dimensions 4 and 6, respectively, then $M \times K$ is a
$\G$-manifold.
\end{cor}

This is rather striking, because string theory focuses attention on
spacetimes of this form, at least if we work in the Euclidean signature.
For example, we can take $M$ to be Euclidean $\R^4$ and $K$ to be any
6-dimensional Calabi--Yau manifold.  Moreover, our remarks on the $\SO(10)$
grand unified theory imply:

\begin{prop} \et Suppose $X$ is a $\G$-manifold. 
Then $X$ has a natural
spin structure, and Dirac spinors at any point of $X$ form a
representation of $\G$ equivalent to the representation of this group 
on the fermions and antifermions in one generation of the Standard Model.
\end{prop}

It would be nice to find a use for these results.

\end{document}